\documentclass[conference]{IEEEtran}
\IEEEoverridecommandlockouts
\usepackage{cite}
\usepackage{amsmath,amssymb,amsfonts}
\usepackage{algorithmic}
\usepackage{graphicx}
\usepackage{textcomp}
\usepackage{xcolor}
\usepackage{hyperref}
\usepackage[latin1]{inputenc}
\def\BibTeX{{\rm B\kern-.05em{\sc i\kern-.025em b}\kern-.08em
    T\kern-.1667em\lower.7ex\hbox{E}\kern-.125emX}}
\usepackage{cleveref}
\usepackage[inline]{enumitem}
\usepackage{xcolor}
\usepackage[nolist]{acronym}
\usepackage{soul}
\usepackage{tikz}

\crefname{section}{Sec.}{Secs.}
\crefname{figure}{Fig.}{Figs.}
\crefname{equation}{Eq.}{Eqs.}

\newenvironment{myitemlist}%
{%
\begin{itemize}[parsep=0em,leftmargin=*,label={--}]%
}%
{%
\end{itemize}%
}

\newcommand{\myfigeps}[3][width=\columnwidth]{%
\begin{figure}[tbp]%
\centering%
\includegraphics[#1]{figures/#2}%
\caption{#3}%
\vspace{-1em}%
\label{fig:#2}%
\end{figure}%
}

\newcommand{\myfigdoubleeps}[5][width=\columnwidth]{%
\begin{figure*}[tbp]%
\centering%
\includegraphics[#1]{figures/#2}%
\includegraphics[#1]{figures/#3}%
\caption{#5}%
\label{fig:#4}%
\end{figure*}%
}

{%
\begin{enumerate*}[label=(\roman*)]%
}%
{%
\end{enumerate*}%
}

\newenvironment{myenumlist}%
{%
\begin{enumerate}[parsep=0em,leftmargin=*,label=\arabic*.]%
}%
{%
\end{enumerate}%
}

\usepackage[separate-uncertainty=true, multi-part-units=single]{siunitx}
\DeclareSIUnit{\bit}{b}
\DeclareSIUnit{\byte}{B}

\usepackage{url}
\makeatletter
\g@addto@macro{\UrlBreaks}{\UrlOrds}
\makeatother

\usepackage{comment}

\usepackage{balance}

\begin{document}

\begin{acronym}
  \acro{3GPP}{Third Generation Partnership Project}
  \acro{5G-PPP}{5G Public Private Partnership}
  \acro{AA}{Authentication and Authorization}
  \acro{ADAS}{Advanced Driver-Assistance Systems}
  \acro{API}{Application Programming Interface}
  \acro{AP}{Access Point}
  \acro{AR}{Augmented Reality}
  \acro{BGP}{Border Gateway Protocol}
  \acro{BSP}{Bulk Synchronous Parallel}
  \acro{BS}{Base Station}
  \acro{CDF}{Cumulative Distribution Function}
  \acro{CFS}{Customer Facing Service}
  \acro{CPU}{Central Processing Unit}
  \acro{DHT}{Distributed Hash Table}
  \acro{DNS}{Domain Name System}
  \acro{ETSI}{European Telecommunications Standards Institute}
  \acro{FCFS}{First Come First Serve}
  \acro{FSM}{Finite State Machine}
  \acro{FaaS}{Function as a Service}
  \acro{GPU}{Graphics Processing Unit}
  \acro{HTML}{HyperText Markup Language}
  \acro{HTTP}{Hyper-Text Transfer Protocol}
  \acro{ICN}{Information-Centric Networking}
  \acro{IETF}{Internet Engineering Task Force}
  \acro{IIoT}{Industrial Internet of Things}
  \acro{IPP}{Interrupted Poisson Process}
  \acro{IP}{Internet Protocol}
  \acro{ISG}{Industry Specification Group}
  \acro{ITS}{Intelligent Transportation System}
  \acro{ITU}{International Telecommunication Union}
  \acro{IT}{Information Technology}
  \acro{IaaS}{Infrastructure as a Service}
  \acro{IoT}{Internet of Things}
  \acro{JSON}{JavaScript Object Notation}
  \acro{LCM}{Life Cycle Management}
  \acro{LL}{Link Layer}
  \acro{LTE}{Long Term Evolution}
  \acro{MAC}{Medium Access Layer}
  \acro{MBWA}{Mobile Broadband Wireless Access}
  \acro{MCC}{Mobile Cloud Computing}
  \acro{MEC}{Multi-access Edge Computing}
  \acro{MEH}{Mobile Edge Host}
  \acro{MEPM}{Mobile Edge Platform Manager}
  \acro{MEP}{Mobile Edge Platform}
  \acro{ME}{Mobile Edge}
  \acro{ML}{Machine Learning}
  \acro{MNO}{Mobile Network Operator}
  \acro{NAT}{Network Address Translation}
  \acro{NFV}{Network Function Virtualization}
  \acro{NFaaS}{Named Function as a Service}
  \acro{OSPF}{Open Shortest Path First}
  \acro{OSS}{Operations Support System}
  \acro{OS}{Operating System}
  \acro{OWC}{OpenWhisk Controller}
  \acro{PMF}{Probability Mass Function}
  \acro{PU}{Processing Unit}
  \acro{PaaS}{Platform as a Service}
  \acro{PoA}{Point of Attachment}
  \acro{QoE}{Quality of Experience}
  \acro{QoS}{Quality of Service}
  \acro{RPC}{Remote Procedure Call}
  \acro{RR}{Round Robin}
  \acro{RSU}{Road Side Unit}
  \acro{SAN}{Storage Area Network}
  \acro{SBC}{Single-Board Computer}
  \acro{SDN}{Software Defined Networking}
  \acro{SDK}{Software Development Kit}
  \acro{SLA}{Service Level Agreement}
  \acro{SMP}{Symmetric Multiprocessing}
  \acro{SRPT}{Shortest Remaining Processing Time}
  \acro{STL}{Standard Template Library}
  \acro{SaaS}{Software as a Service}
  \acro{TCP}{Transmission Control Protocol}
  \acro{TSN}{Time-Sensitive Networking}
  \acro{UDP}{User Datagram Protocol}
  \acro{UE}{User Equipment}
  \acro{URI}{Uniform Resource Identifier}
  \acro{URL}{Uniform Resource Locator}
  \acro{UT}{User Terminal}
  \acro{VANET}{Vehicular Ad-hoc Network}
  \acro{VIM}{Virtual Infrastructure Manager}
  \acro{VM}{Virtual Machine}
  \acro{VNF}{Virtual Network Function}
  \acro{VR}{Virtual Reality}
  \acro{WLAN}{Wireless Local Area Network}
  \acro{WMN}{Wireless Mesh Network}
  \acro{WRR}{Weighted Round Robin}
  \acro{YAML}{YAML Ain't Markup Language}
\end{acronym}

\title{Stateless or stateful FaaS? I'll take both!
}

\author{%
\IEEEauthorblockN{Carlo Puliafito\textsuperscript{\textsection}}
\IEEEauthorblockA{\textit{University of Pisa} \\
}
\and
\IEEEauthorblockN{Claudio Cicconetti\textsuperscript{\textsection}}
\IEEEauthorblockA{\textit{IIT-CNR} \\
}
\and
\IEEEauthorblockN{Marco Conti}
\IEEEauthorblockA{\textit{IIT-CNR} \\
}
\and
\IEEEauthorblockN{Enzo Mingozzi}
\IEEEauthorblockA{\textit{University of Pisa} \\
}
\and
\IEEEauthorblockN{Andrea Passarella}
\IEEEauthorblockA{\textit{IIT-CNR} \\
}
}


\maketitle

\begingroup\renewcommand\thefootnote{\textsection}
\footnotetext{C.~Puliafito and C.~Cicconetti share the first author role in this paper.\\
\\
    \textcopyright 2022 IEEE.  Personal use of this material is permitted.  Permission from IEEE must be obtained for all other uses, in any current or future media, including reprinting/republishing this material for advertising or promotional purposes, creating new collective works, for resale or redistribution to servers or lists, or reuse of any copyrighted component of this work in other works.
}
\endgroup

\begin{abstract}
Serverless computing has emerged as a very popular cloud technology, together with its companion Function-as-a-Service (FaaS) programming model enabling invocations of stateless functions from clients.
An evolution of serverless is now taking place, shifting it towards the edge of the network and broadening its scope to stateful functions, as well.
In this paper we argue that stateless vs.\ stateful is not a dichotomy of the application \textit{per se}, but rather a time-varying property of most (if not all) applications, as confirmed by the analysis of real traces collected in a production environment.
Based on this observation, we propose a mathematical formulation of a resource allocation problem that jointly encompasses both operation modes, dubbed \emph{lambda} vs.\ \emph{mu}, which can be solved efficiently at run-time by an edge orchestrator.
We evaluate the proposed solution via simulation experiments in realistic network and workload conditions, which leads the way to the practical realization of a system where applications can freely adapt their current operation mode and optimize their performance at a minimum cost of operation from the network's perspective.
\end{abstract}


\section{Introduction}\label{sec:intro}

Edge computing is a powerful extension of cloud computing toward the network edge.
It consists of geographically distributed compute nodes located in proximity to access networks (far-edge nodes) or within the core network of the telco operator (near-edge nodes)~\cite{edgesurvey}. Edge nodes run \textit{microservices}: small pieces of code that are often packaged inside containers rather than virtual machines because they are faster to boot up and more lightweight~\cite{MorabitoContvm}.
%
We can identify two main ways to operate microservices and realize an end user application: \ac{FaaS} vs.\ \ac{PaaS}.

\ac{FaaS} was initially designed for cloud data centers~\cite{Vaneyk2019} but is rapidly gaining momentum in edge computing, too~\cite{Xie2021}. With FaaS, a microservice (called \textit{function}) can be instantiated in several containers that are equivalent to one another and, hence, can be autoscaled by the platform provider with maximum flexibility. Such an equivalence allows consecutive invocations from the same client to be forwarded to different containers, and a given container to serve multiple clients.
Besides, \ac{FaaS} enables a pure pay-per-use model, where billing is based on the number of function invocations or cumulative execution time, regardless of the rate of invocations.
One disadvantage of \ac{FaaS} is that the containers cannot keep any state associated to the application's session~\cite{Eismann2021a}:
every time a function is invoked, if needed, it must read (write) the session state from (to) a remote storage service (e.g., located in the cloud), which increases latency and incurs extra costs. From now on, we refer to \ac{FaaS} containers as \textit{stateless}.


On the other hand, with \ac{PaaS} a container is dedicated to a user application instance so that: i) all the invocations from the client are forwarded to that container; and ii) that container handles invocations from that client only. Since the container is dedicated to the client, it keeps the session state locally, hence we call it \textit{stateful}. In contrast to the previous approach, this one reduces latency, as session state does not need to be accessed from a remote storage. As a result, this approach is widely used by edge platforms, and also FaaS platforms for edge computing are starting to consider stateful containers as a possible alternative to stateless~\cite{Lopez2021}. Yet, this approach falls short of flexibility and cost-efficiency: in general, a dedicated container is expensive for the user (especially at the edge) as resources are paid for the whole time during which the application is active, which is inefficient with a sporadic use.

In the literature and market technology, the stateless and stateful operation modes are considered as alternatives, with the choice being made by the developer at design time.
However, it can happen that the very same application has a heterogeneous usage pattern over time, which may result in degraded performance (during peaks under a stateless approach) or wasted resources (during sporadic use under a stateful approach).
Therefore, in this work we get a new perspective, and propose instead to let an application \textit{adapt dynamically} to the best operation mode, i.e., to switch from being stateless to stateful, and \textit{vice versa}, depending on the current conditions.
%
%
The contribution of this work is threefold:
 \begin{myitemlist}
    \item we show with a quantitative analysis of public traces obtained in the wild that alternating between operation modes minimizes the cost of operation, in terms of the container renting fees (stateful), function invocations and storage services (stateless), and migration overhead (\cref{sec:motivation});
    %
    %
    \item we formulate a problem that jointly optimizes the placement of stateful containers and the distribution of function invocations to stateless containers at the edge, and propose an efficient solution and practical implementation (\cref{sec:model});
    \item we evaluate the performance of the proposed system through simulations under realistic network and workload conditions, to identify the key trade-offs incurred by the configuration of the system parameters (\cref{sec:eval}).
 \end{myitemlist}
The paper also includes \cref{sec:soa} to position our work in the state-of-the-art and \cref{sec:conclusions}, which concludes the paper and outlines the future work.
 

\section{Related work}\label{sec:soa}



Many big players offer FaaS solutions to their customers, such as Amazon with AWS Lambda, Microsoft with Azure Functions, and IBM with Cloud Functions, just to name a few. Although these systems were initially designed for cloud environments, there are now extensions toward the network edge: Amazon Lambda@Edge, Microsoft Azure Edge Zones, and IBM Edge Functions. All the above platforms adhere to the typical FaaS approach where functions are served as stateless containers. However, we highlight that stateful containers are gradually coming into the picture as a complementary approach. Specifically, Microsoft introduces the concept of \textit{entity functions}~\cite{MicrosoftDurable}, which are uniquely identified, dedicated resources that keep the session state locally as an in-memory object. \textit{Long-lived functions}~\cite{amazonlonglived} from Amazon and \textit{Durable objects}~\cite{cloudflaredurable} from Cloudflare are other examples of dedicated resources from commercial FaaS platforms.

Besides platforms from companies, some open-source FaaS solutions are also available, e.g., Apache OpenWhisk, OpenFaaS, Kubeless, and Knative. All of them leverage Kubernetes as orchestration system underneath. In Kubernetes, function instances are called Pods, which can encapsulate one or more containers. Kubernetes defines both stateless and stateful Pods, the latter being implemented by matching persistent volumes to uniquely identified Pods~\cite{statefulpod}.

In the scientific domain, there are some works in the direction of realizing a coexistence of stateless and stateful containers in FaaS systems, especially from the point of view of the programming model to be used and related \acp{API}.
For instance, Baresi~\textit{et al.}~\cite{baresi} describe the proof-of-concept implementation of a FaaS platform for edge computing, based on Apache OpenWhisk, also mentioning stateful containers for uniquely identified resources.
\textit{However, none of the works so far consider the possibility for a function to dynamically adapt its operation mode over time, which we have hinted in our previous work~\cite{statefulfaasieeecomputer} and investigate in detail here.}


On the other hand, a well-studied topic in edge computing is the optimal placement of dedicated microservices in the infrastructure.
It is known that algorithms that are widely used in cloud data centers cannot be exploited \textit{as-is} at the edge, due to the distinctive characteristics of this environment, e.g., wide-area deployment and resource limitations of edge nodes.
%
%
As comprehensively described in related surveys~\cite{sonkolysurvey2021, salahtsurveyacm}, most of the scientific works formalize the problem as a linear programming one where the objective function typically aims at optimizing latency, energy, or resource utilization. As optimization constraints, authors usually consider network limitations (e.g., bandwidth capacity or network latency) and compute ones (e.g., available processing power and memory).

Given the relatively newer topic, fewer works instead aim at optimizing the distribution of function invocations to stateless containers. The work in~\cite{cicconettitnsm2021} proposes a decentralized framework where entry points to the system take autonomous decisions on where to forward function invocations, based on weights that are dynamically and locally updated to minimize the communication latency. In~\cite{Mittal2021}, function invocations are dispatched based on the queue length and service capacity of each container, with the aim to minimize latency.
\textit{To the best of our knowledge, there are no works formulating an optimization problem that jointly aims at optimizing placement of stateful containers and dispatching of invocations to stateless containers, which we address in \cref{sec:model}.}

\section{Motivation}\label{sec:motivation}

In this section, we report the findings of our analysis of real \ac{FaaS} traces collected in a period of two weeks in 2020 on Microsoft Azure Functions and made available in a public dataset\footnote{\url{https://github.com/Azure/AzurePublicDataset/blob/master/AzureFunctionsBlobDataset2020.md}}, thoroughly analyzed in \cite{Romero2021FaaT:Applications}.
The dataset contains more than 44 millions of anonymized function invocations from 856 applications. For each invocation a set of data are included, from which we use the following: the timestamp, unique identifiers of the user ID and application name, and a flag specifying whether the application's state has been accessed in read or write mode.
The applications sampled in the dataset are very heterogeneous, e.g., the number of daily invocations ranges from very few to millions.
Read accesses are 77\% of the total.

\textit{Our objective is to show that the majority of those applications can benefit from a policy that adapts their stateful vs.\ stateless nature over time, in terms of some performance metric, which in the following we assume to be the cost of operation under some reasonable simplifying assumptions.}
In particular, we assume that the cost of a stateful application is given only by the duration of the time window when it is assigned a dedicated container:

\begin{equation}\label{cost-mu}
    c_\mu = \Omega_\mu T_\mu,
\end{equation}
where $\Omega_\mu$ is the cost per time unit and $T_\mu$ is the time units the application spent as stateful.
On the other hand, for a stateless application we assume that its cost is given by the number of invocations and the type of state access, as follows:

\begin{equation}\label{cost-lambda}
    c_\lambda = \xi_\lambda \left( N_\lambda^R + N_\lambda^W \right) + \sigma_\lambda^R N_\lambda^R + \sigma_\lambda^W N_\lambda^W,
\end{equation}
where $\xi_\lambda$ is the cost per function invocation, $\sigma_\lambda^R$ ($\sigma_\lambda^W$) is the cost per read (write) access, and $N_\lambda^R$ ($N_\lambda^W$) is the number of function invocations with read (write) accesses.

\myfigdoubleeps{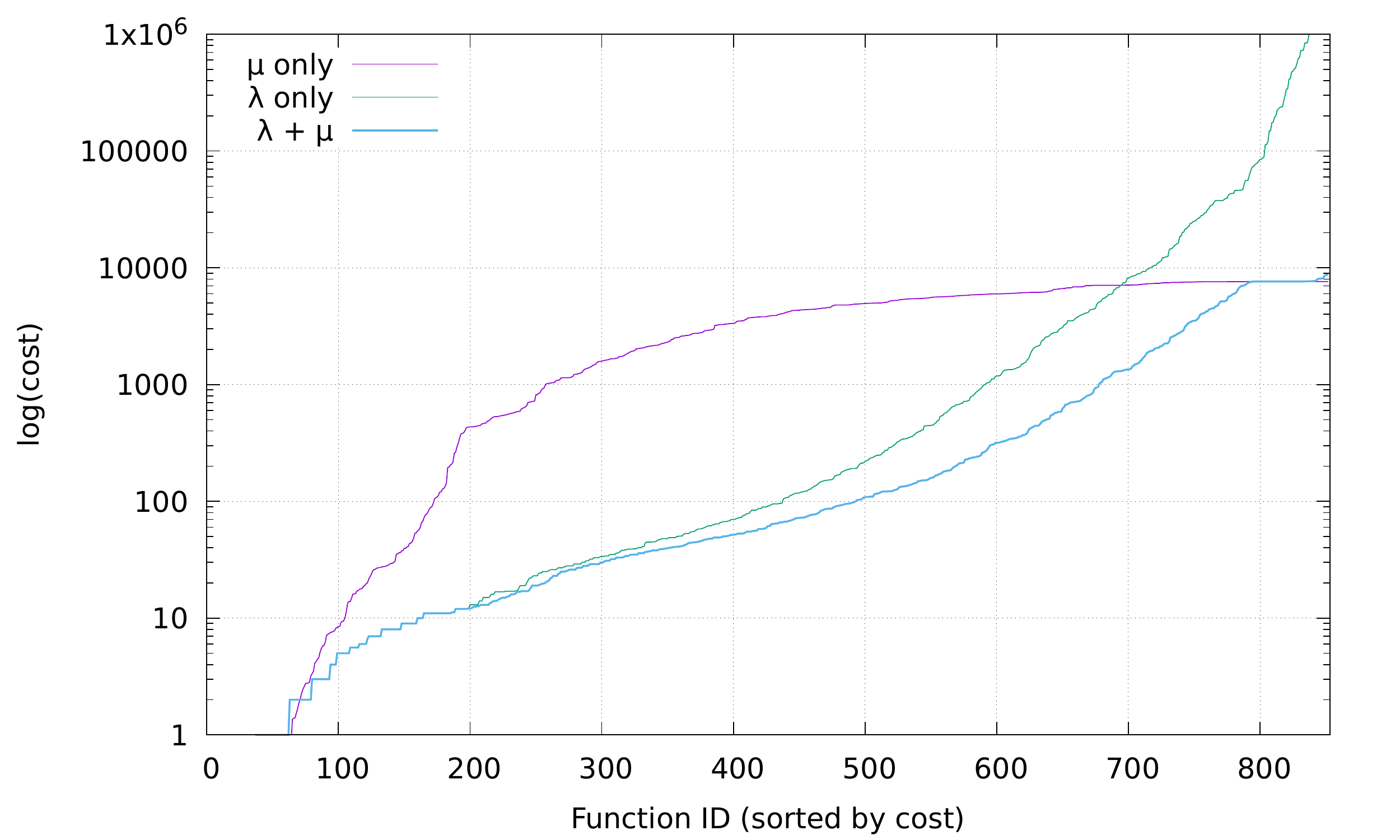}{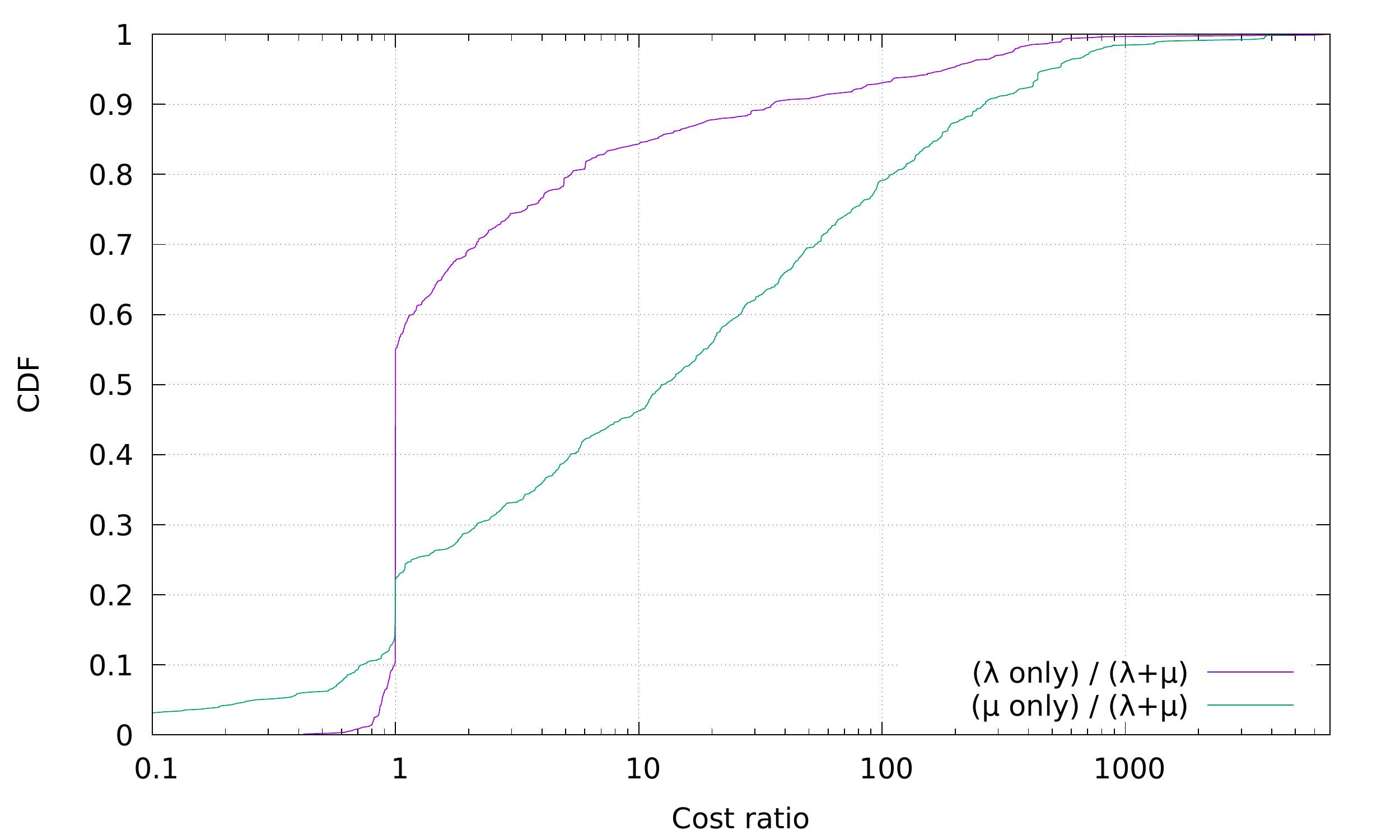}{cost-comparison}{%
Comparison of the cost of execution of the applications in the Microsoft Azure Database~\cite{Romero2021FaaT:Applications} with $\lambda$-only vs.\ $\mu$-only vs.\ $\lambda+\mu$ policies, with $\xi_\lambda = 0.6$, $\sigma_\lambda^R = 0.4$, $\sigma_\lambda^W = 5$, $\tau_\lambda = \tau_\mu = 12$, and $\Omega_\mu = 6.3 \cdot 10^{-6}$: absolute (left) and relative (right). All costs in $10^{-6}~\$.$
}

Computing the cost of an application in the dataset with $\lambda$-only and $\mu$-only policies is straightforward.
For the hybrid case, called $\lambda+\mu$, where an application migrates from stateful to stateless, we have defined two migration costs ($\tau_{lambda}$: from stateful to stateless; $\tau_{mu}$: from stateless to stateful) and implemented the following policy:

\begin{myitemlist}
\item if an application is currently run as stateful, it migrates to stateless if keeping the container occupied until the next function invocation, which costs $\Omega_\mu \left( t_{\mathrm{next}} - t_{\mathrm{now}} \right)$, is more expensive than migrating to stateless right now;
\item otherwise, if an application is currently run stateless, we perform a simulation in a look-ahead window of future call invocations in the cases migration-to-$\mu$ vs.\ keep-as-$\lambda$; we then migrate it to stateful if a break-even point is reached.
\end{myitemlist}

Note that both policies are heuristic but require prophetic powers to predict the precise future pattern of function invocations, which is almost always not available to the platform or the application logic programmer as it depends on external circumstances.
However, this assumption is consistent with our goal of showing that a suitable policy \textit{exists}, not how it could be realized effectively in real settings.

The values used for the cost model are reported in the caption and they are inspired from publicly available prices of Amazon Lambda@Edge\footnote{\url{https://aws.amazon.com/lambda/pricing/?nc1=h_ls}}, where (e.g.) the invocation of 1 million functions costs \$0.6, and the cost of a \texttt{GET} (\texttt{PUT}) operation to read (write) the state is about \$0.4 (\$5) for 1 million operations.
In the absence of a more realistic model, the migration cost in either direction has been arbitrarily estimated as twice the cost of function invocation + read + write.
The figures reported are purely indicative, e.g., they do not include storage costs and they do not take into account volume discounts, and subject to change depending on the region, provider, as well as to adapt to the evolution of technology and business models.
However, we believe these simplified assumptions are sufficient for our purposes.
We show in \cref{fig:cost-comparison} the costs obtained with the three policies.
As can be seen from the left part of the figure, showing the absolute costs, the $\mu$-only and $\lambda$-only curves intersect: some applications are better served \textit{always} as stateful while others as stateless, the latter being the majority in the dataset used with the cost model values adopted.

\noindent\underline{Key observation.} \textit{However, by using a $\lambda+\mu$ hybrid policy, the cost can be minimized for all functions, which confirms our intuition that all applications should be able to alternate between stateful and stateless in their lifetime.}

The relative advantage, in terms of cost, of $\lambda+\mu$ compared to $\lambda$-only and $\mu$-only, respectively, is shown in the right hand side of \cref{fig:cost-comparison}: most of the applications have a cost ratio $> 1$, which becomes substantial for a significant fraction of them, especially in the $\mu$-only case.
We note that for very few applications the cost ratio is $< 1$: this happens because of edge effects of the analysis and only for applications that \textit{absolutely always} are required to remain as either stateful or stateless to minimize their cost.
We have decided not to prune the dataset from such applications, for better transparency of the analysis, but such applications have negligible statistical significance, and they are anyway of little interest for our work.



The tool source code and scripts for this cost analysis on the Azure dataset are publicly available on GitHub\footnote{\url{https://github.com/ccicconetti/support}, tag \texttt{dataset-001}, check out the instructions in \texttt{Dataset/001\_Mu\_Lambda/README.md}.}.

\section{System Model}\label{sec:model}

\begin{figure}[t!]
 \centering
 \includegraphics[width=.85\columnwidth]{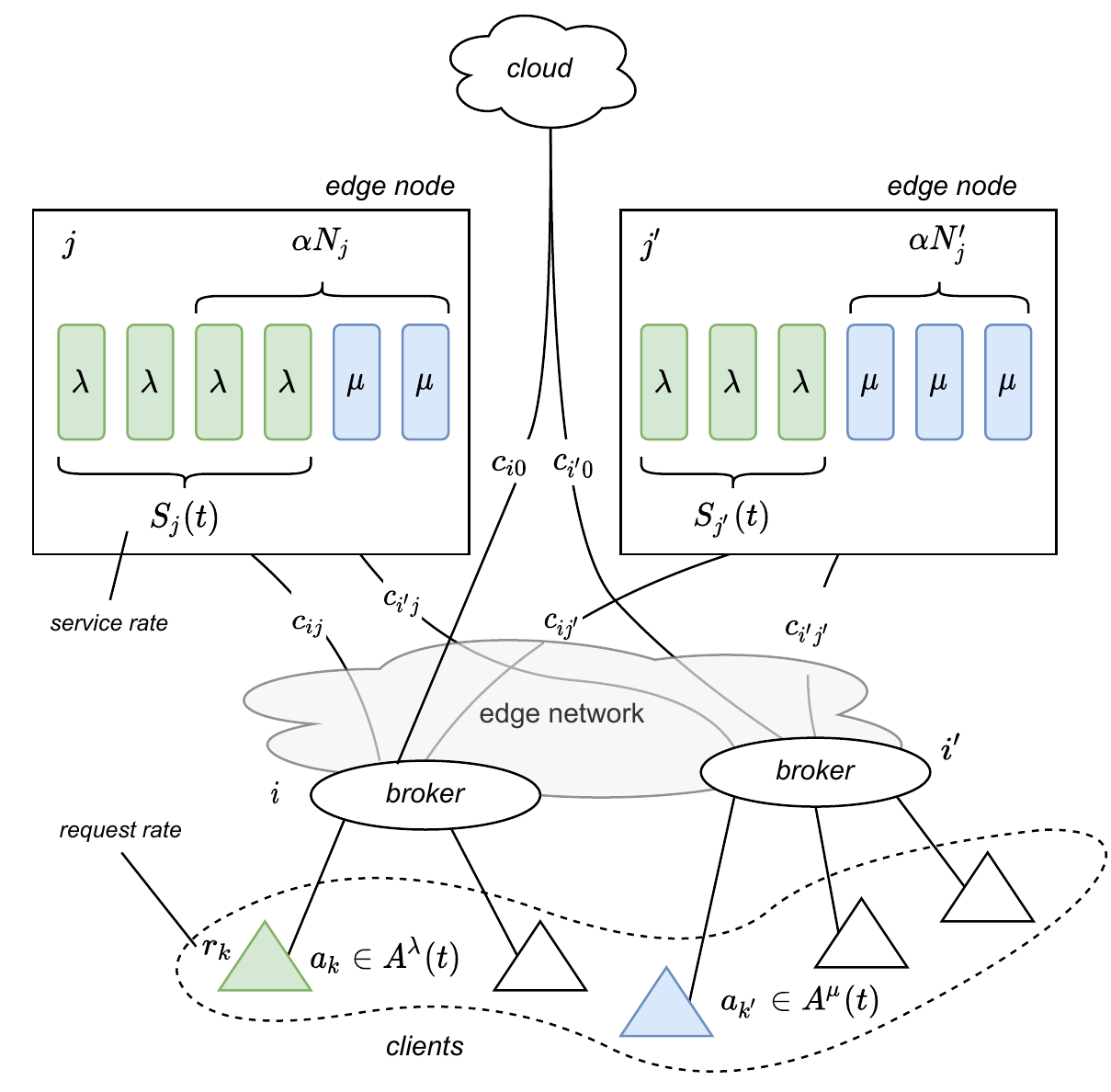}
 \caption{System model and notation.}\label{fig:model-1}
\end{figure}

Our system is modeled as follows and illustrated in \cref{fig:model-1}.
We have a set of \textit{clients} that use services provided by edge or cloud nodes, which are
reached through \textit{brokers} located at the network edge that represent entry points to the system.
We assume for simplicity of notation that each client hosts a single application and we only consider those that are alive and active.
The \textit{cloud} resources are assumed to be unlimited, while \textit{edge nodes} have a finite number of containers reserved for the service, but as all the clients are located at the edge of the network it is always ``cheaper'' to run applications on them compared to the cloud.
Such a cost could refer to the use of network resources (point of view of the edge infrastructure operator) or to the latency (point of view of the end users).
%
Note that the ``cost'' in this section is different from that in \cref{sec:motivation}: the latter is assumed to be minimized by applications by switching back and forth between the $\mu$ vs.\ $\lambda$ modes of operation to adapt to a changing environment; instead, in the following we adopt the perspective of the infrastructure operator and strive to minimize the \textit{operational costs}.


At a given time, as already discussed, an application can be in one of two possible states depending on its internal operation and environmental conditions: i) \textit{stateless} (we call it a $\lambda$-app), where a pool of containers is shared among a set of applications invoking stateless functions, vs.\ ii) \textit{stateful} (we call it $\mu$-app), where the application invokes stateful function calls, which require persistence on a dedicated containerized microservice.
%
%
We assume that the transition from one state to another is mediated in the edge domain by an orchestrator (not shown in \cref{fig:model-1}), which is in charge of: i) handling transition requests from the applications, ii) deciding whether to assign a container of a $\mu$-app to the cloud or to the edge, and in the latter case on which edge node, and iii) configuring the brokers so that the stateless functions invocations can be dispatched to the containers shared by $\lambda$-apps.
In this work, we focus on the decision process for resource allocation of $\mu$- and $\lambda$-apps, which we model mathematically in \cref{sec:model:problem}, and for which we provide a solution and an implementation scheme, respectively in \cref{sec:model:solution} and \cref{sec:model:operation}.
We do not elaborate on the protocols and interfaces that would be needed for the practical deployment, which is left for future work.

\subsection{Problem formulation}\label{sec:model:problem}

We now formally define the resource allocation problem, taking into account jointly the $\mu$- and $\lambda$-apps.
Again, with reference to  \cref{fig:model-1}, let $A = \{a_k\}$ be the set of application clients and $B = \{b_i\}$  the set of brokers.
We define the association $y_{ki}$ between a client $a_k$ and a broker $b_i$ as follows:

\begin{equation}\label{eq:yki_definition}
 y_{ki}=\begin{cases}
    1, & \text{if $a_k$ is bound to $b_i$}\\
    0, & \text{otherwise}
  \end{cases}
\end{equation}

Besides, $y_{ki}$ has the following property:
\begin{equation}\label{eq:yki_property}
    \sum_{i} y_{ki} = 1, \forall k
\end{equation}
which states that each client is bound to one broker only. A broker $b_i$ receives function invocations from its clients and dispatches such invocations to containers, which are function instances running on compute nodes. In our system, $E = \{e_j\}$ is the set of compute nodes. Each compute node $e_j$ is assumed to be deployed at the network edge (i.e., edge nodes) and to have $N_j$ containers instantiated on it. The only exception is represented by $e_0$, which is a cloud node having $N_0 = |A|$ containers instantiated on it.

Moreover, we indicate with $A^\lambda(t)$ and $A^\mu(t)$ the subset of clients requiring at time $t$ to be served by $\lambda$- and $\mu$-containers, respectively. At any time $t$, it occurs that $A^\lambda(t)$ $\cup$ $A^\mu(t)$ = $A$ and $A^\lambda(t)$ $\cap$ $A^\mu(t)$ = $\emptyset$.
We also define $x_{kj}(t)$ as follows:

\begin{equation}\label{eq:xkj_definition}
 x_{kj}(t)=\begin{cases}
    1, & \text{if $a_k$ has a $\mu$-container on $e_j$ at time $t$}\\
    0, & \text{otherwise}
  \end{cases}
\end{equation}

At any time $t$, $x_{kj}(t)$ is subject to the following two constraints:

\begin{equation}\label{eq:xkj_constraintone}
    \sum_{j} x_{kj}(t) = 1, \forall k \in A^\mu(t)
\end{equation}

\begin{equation}\label{eq:xkj_constraintthree}
    x_{kj}(t) = 0, \forall k \in A^\lambda(t)
\end{equation}

Eq. (\ref{eq:xkj_constraintone}) guarantees that each client requiring a dedicated container at time $t$ receives exactly one. Eq. (\ref{eq:xkj_constraintthree}) instead states that clients requiring at time $t$ to be served by $\lambda$-containers cannot be given at the same time a dedicated $\mu$-container.

To guarantee that, at any time $t$, enough resources are available to $\lambda$-containers on any node $e_j$, we define a further constraint as follows:
\begin{equation}\label{eq:wjmu_limit}
    \sum_{k} x_{kj}(t) \leq \alpha \cdot N_j,
\end{equation}
where $0 \leq \alpha \leq 1$. The above constraint states that the number of $\mu$-containers that are instantiated on node $e_j$ at time $t$ cannot exceed a pre-defined fraction of $N_j$.
Limit cases: $\alpha = 0$ means that $\mu$-containers \textit{cannot} be assigned to edge node $j$; with $\alpha = 1$ all the resources can be used by $\mu$-containers.



For what concerns clients $a_k \in A^\lambda(t)$, we define $r_k$ as their request rate, i.e., the rate at which those clients invoke $\lambda$-containers. Therefore, we can define the request rate exiting any broker $b_i$ at time $t$ as:
\begin{equation}\label{eq:requestrateoutofbroker}
    R_i(t) = \sum_{k \in A^\lambda(t)} y_{ki} \cdot r_k
\end{equation}

In a similar way, $s_j$ indicates the service rate of a $\lambda$-container running on $e_j$, namely the rate at which that type of container can serve invocations. Given that the cloud node is considered to have unlimited resources, we set $s_0 > \max_{k} \{r_k\}$. We define the available service rate at time $t$ of any node $e_j$ as:

\begin{equation}\label{eq:servicerateedgenode}
    S_j(t) = s_j \cdot \sum_{k} (1 - x_{kj}(t)),
\end{equation}
which is an aggregate of the service rates of all the containers in $e_j$ that are not assigned to $\mu$-apps.

Any broker $b_i$ dispatches invocations to $\lambda$-containers by distributing such invocations toward compute nodes, based on weights $w_{ij}(t)$:
over a sufficiently large time horizon, the ratio between the function invocations dispatched by the broker $i$ toward the edge nodes $1$ and $2$ will be $w_{i1}/w_{i2}$.
At any time $t$, these weights are subject to the following three constraints:
\begin{equation}\label{eq:weightconstraintone}
    w_{ij}(t) \geq 0, \forall i, \forall j 
\end{equation}
\begin{equation}\label{eq:weightconstrainttwo}
    \sum_{j} w_{ij}(t) = 1, \forall i
\end{equation}
\begin{equation}\label{eq:weightconstraintthree}
    \sum_{i} w_{ij}(t) \cdot R_i(t) \leq \beta \cdot S_j(t), \forall j,
\end{equation}
where $0 < \beta < 1$. Specifically, constraint (\ref{eq:weightconstraintthree}) ensures stability by stating that at any time $t$ the request rate entering any node $e_j$ cannot exceed a fraction of the available service rate of that node. Parameter $\beta$ is introduced to allow for some service capacity over-provisioning.

Finally, we define $c_{ij} > 0$ as a cost over the path interconnecting $b_i$ and $e_j$. Following the considerations made at the beginning of \cref{sec:model}, this cost could be related to the usage of network resources, to the communication latency, or to a combination of both. Note that for any broker $b_i$, we set $c_{i0} > \max_{j} \{c_{ij}\}$, which means that reaching the cloud node is always more expensive than reaching any edge node.

Given the above definitions and constraints, we formulate the following optimization problem:

\begin{equation}\label{eq:optimization_both}
    \min_{x_{kj}(t), w_{ij}(t)} \Big\{\Omega
        \sum_{k,i,j} c_{ij} \cdot x_{kj}(t) +
        \sum_{i,j} c_{ij} \cdot w_{ij}(t) \cdot R_i(t)
        \Big\},
\end{equation}
where $\Omega$ is big enough that the first term always dominates over the second one.
The above problem aims at instantiating $\mu$-containers on compute nodes and finding the weights $w_{ij}(t)$ that allow to dispatch invocations to $\lambda$-containers so as to minimize a combined overall cost in the system.

\noindent\underline{Key observation.} \textit{The objective function in \cref{eq:optimization_both} stipulates that the use of edge resources is preferred for $\mu$-applications, which is counterbalanced by the selection of a minimum amount of containers $(1 - \alpha) N_j$ reserved for $\lambda$-applications in each edge node $e_j$.}

\subsection{Solution}\label{sec:model:solution}

The constraints \cref{eq:yki_definition}-\cref{eq:weightconstraintthree} and the objective function \cref{eq:optimization_both} form a mixed integer linear programming problem, as the variables $x_{ky}(t)$ (integer)  and $w_{ij}(t)$ (real) only exhibit linear relationships.
Furthermore, thanks to our assumption that $\Omega \gg 1$, it is possible to separate the problem into two sub-problems, which can be solved sequentially and still achieve the global optimum, as given by the following objective functions with the following procedure at time $t$:

\begin{myenumlist}
    \item \ul{$\mu$-apps allocation sub-problem}: find $x_{kj}(t)$ with objective function \cref{eq:optimization_mu}:
    \begin{equation}\label{eq:optimization_mu}
        \min_{x_{kj}(t)} \sum_{k,i,j} c_{ij} \cdot x_{kj}(t),
    \end{equation}
    which means that all the $\mu$-apps will be assigned to a container on the edge nodes (or in the cloud).
    As a result of the allocation in the previous step, all the containers for which it is $x_{kj}(t) = 1$ will not contribute to the execution of $\lambda$-app function invocations, as $(1-x_{kj}(t))$ will be $0$ in \cref{eq:servicerateedgenode}.
    \item \ul{$\lambda$-apps allocation sub-problem}: find $w_{ij}(t)$ with objective function \cref{eq:optimization_lambda}:
    \begin{equation}\label{eq:optimization_lambda}
        \min_{w_{ij}(t)} \sum_{i,j} c_{ij} \cdot w_{ij}(t) \cdot R_i(t),
    \end{equation}
    which means that load balancing of $\lambda$-apps at each broker $b_i$ will happen in accordance with the weights found; we recall that stability is ensured by \cref{eq:weightconstraintthree}.
\end{myenumlist}

Both sub-problems in steps 1 and 2 above are instances of well-known optimization problems. More specifically, the first one is a case of \textit{assignment problem} and the second one of \textit{transportation problem}, and both can be solved (exactly) with efficient algorithms from the operations research literature~\cite{ahuja1993network}.

For example, to carry out the performance evaluation in the next section, we use the following algorithms:
for the $\mu$-apps allocation problem we adopt the Hungarian method, which has $\mathcal{O}(|A^\mu(t)|^3)$ worst-case time complexity; on the other hand, we transform the $\lambda$-apps allocation problem into an equivalent minimum cost flow problem, which we then solve using the ``successive shortest path'', having worst-case time complexity $\mathcal{O}\left(\bar{R} \cdot \left(E + V \log V\right)\right)$, where:
\begin{align*}
    \bar{R} & = \sum_{k \in A^\lambda(t)} r_k, \\
    E       & = \bar{A} \cdot \bar{N} + 2 \cdot (\bar{A} + \bar{N}),\\
    V       & = 2 \cdot (\bar{A}+\bar{C}+1), \\
    \bar{A} & = |A^\lambda(t)|, \\
    \bar{N} & = \sum_{j} \left[ N_j - \sum_{k} x_{kj}(t) \right]. \\
\end{align*}



\subsection{Overall operation}\label{sec:model:operation}

The solution illustrated in the previous section provides the optimal allocation, under the given constraints and costs, for a given set of applications $A^\lambda(t) \cup A^\mu(t)$.
In principle, this implies that whenever any of the following happens, the algorithm has to be re-run:
i) a new application becomes active;
ii) an active application become inactive;
iii) an application migrates from $\mu$-app to $\lambda$-app or \textit{vice versa}.
If the population of users is large or the frequency of changes is high, the allocation will have to be adjusted very often, which in turn has two consequences.
First, the orchestrator may become a performance bottleneck: even though we have formulated the problem so that efficient solutions can be used, finding an exact solution in a short amount of time can be a challenge with large problem instances, which needs to be addressed either by using a big amount of computational resources (costly and with environmental sustainability concerns) or by finding approximate solutions (possibly degrading the performance).
Second, whenever a new allocation of resources is found by the orchestrator, some of the current $\mu$-apps may need to be migrated from one edge node to another, which is undesirable both for the edge infrastructure operator (network resources are consumed) and for the end users (possible service interruptions and latency spikes).

To address this issue we propose to run the resource allocation algorithm in \cref{sec:model:solution} only periodically (we call the period \textit{epoch}).
Of course, asynchronous events may happen in between epochs, which we can handle in a best-effort manner as follows:
\begin{myenumlist}
\item a new $\lambda$-app becomes active: we configure brokers in such a way that invocations of unknown functions are directed automatically to the cloud, hence no reconfiguration is needed at the edge (even though the allocation of $w_{ij}$ is sub-optimal in general);
\item a new $\mu$-app becomes active: the orchestrator assign it to the edge container with minimum cost $c_{ij}$ among those edge nodes $j$ that have available resources for this according to \cref{eq:wjmu_limit}, which is a $\mathcal{O}(|E|)$ operation; if no such edge node is available, then the container is created in the cloud; in any case, no re-allocation of the other currently active $\mu$-apps is done;
\item a $\lambda$-app becomes inactive: no operation needed on the brokers, who will simply not receive anymore function invocations from the corresponding client;
\item a $\mu$-app becomes inactive: the container assigned is deallocated, which frees resources on the corresponding node;
\item migration $\mu \rightarrow \lambda$: execute the actions in point~4 followed by those in point~1;
\item migration $\lambda \rightarrow \mu$: execute the actions in point~2.
\end{myenumlist}
The epoch duration is a system parameter that has to be tuned appropriately since it incurs a performance trade-off, which we study with simulations in \cref{sec:eval}.

\section{Evaluation}\label{sec:eval}

In this section we assess the performance of the framework proposed in \cref{sec:model} using numerical simulations.
For reproducibility purposes, the tool used is released as open source on GitHub\footnote{\url{https://github.com/ccicconetti/serverlessonedge/tree/v1.3.0}, see instructions in \texttt{README.md} within simulations \texttt{010}, \texttt{011}, and \texttt{012}.}, together with the artifacts and the scripts to run the experiments and analyze the output.


\begin{figure}[tbp!]
 \centering
 \includegraphics[width=.85\columnwidth]{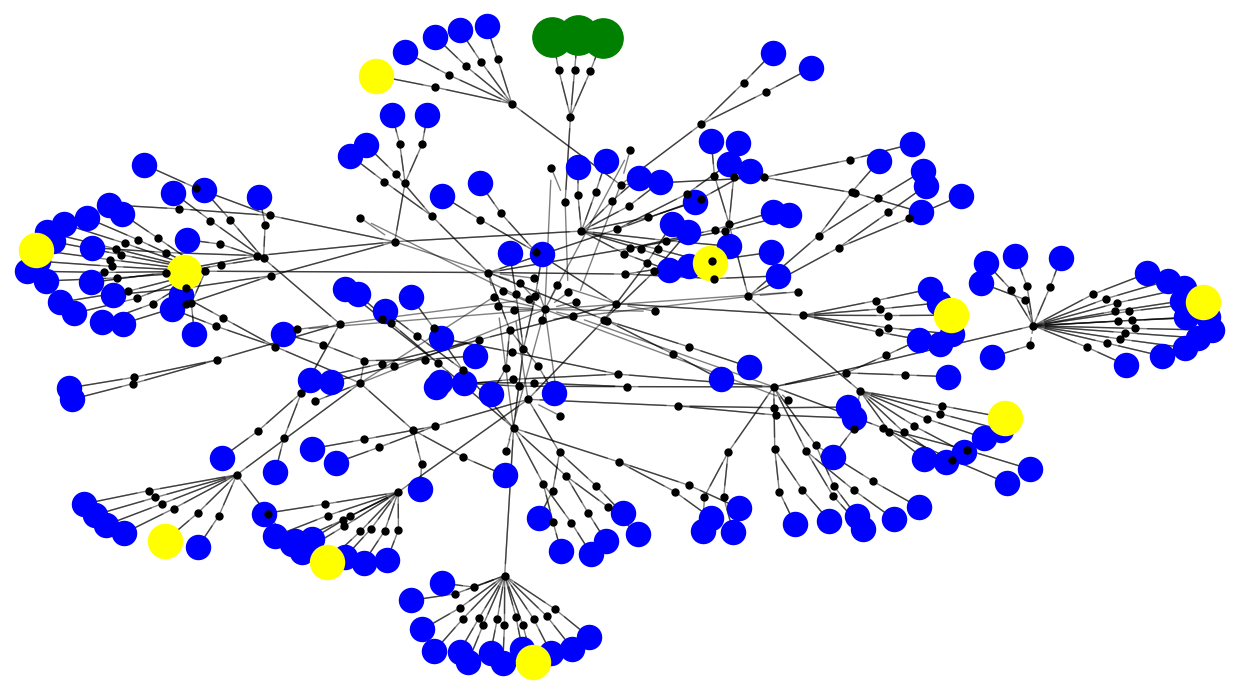}
 \caption{Network topology: blue nodes are the brokers, yellow nodes are far edges, green nodes are near edges, black dots are network devices.}\label{fig:topology}
\end{figure}

The network topology used is depicted in \cref{fig:topology}: it was generated with ``ether: Edge Topology Synthesizer''\footnote{\url{https://github.com/edgerun/ether}.},
which produces realistic edge network models with mixed compute nodes: end-user devices (our brokers, in blue), small PCs (far-edge nodes, in yellow), and servers (near-edge, in green).
We have used as network cost $c_{ij}$ the logical distance between broker $i$ and edge node $j$, in number of hops.
The cloud node is not included in the topology but it is considered in the resource allocation with cost $c_{i0} = 2 \cdot \max\left\{c_{ij}\right\}$.
The request rate of $\lambda$-apps is assumed to be $r_i = 1$ for all applications, while the service rate is $s_j = 10$ for far-edge nodes and $s_j = 20$ for near-edge nodes; similarly, the number of containers available in each node is $N_j = 4$ in far-edge devices and $N_j = 8$ in near-edge devices.
In the following we report the results obtained with two types of simulation: snapshot and dynamic.

\subsection{Snapshot simulations}\label{sec:eval:snapshot}

Snapshot simulations follow a Monte Carlo approach: a number of $\lambda$-apps ($\mu$-apps) is drawn from a Poisson distribution with mean $E[|A^\lambda|]$ ($E[|A^\mu|]$); each app is assigned to a random broker selected independently from those available with uniform distribution probability; we execute the algorithms in \cref{sec:model:solution} and find the costs of $\lambda$- and $\mu$-apps.
We then repeat the same process for several replications (6400 in our experiments), all contributing to the same experiment instance.
We have run instances with $E[|A^\lambda|] = 50$ while changing $\alpha \in \{ 0/8, 1/8, \ldots, 7/8 \}$, $\beta \in \{ 0.1, \ldots, 0.9 \}$, and $E[|A^\mu|] \in \{ 25, 50, 75 \}$ in a full combinatorial manner.


In \cref{fig:012-lambda-cost} we study the effect of $\alpha$ and $\beta$ on the (unitary) cost of $\lambda$-apps, defined as the value of the summation in \cref{eq:optimization_lambda} divided by the number of $\lambda$-apps in the snapshot.
As the fraction of containers per edge node reserved to $\lambda$-apps is $(1-\alpha)$, the  cost increases with $\alpha$.
%
The impact of $\alpha$ is much more prominent with small values of $\beta$, when the cost is higher.
In fact, $\beta$ controls how much margin the orchestrator should reserve to cope with deviations of the actual rate of requests from applications compared to the nominal values $s_i$ provided to the algorithm: e.g., $\beta = 0.1$ means that we consider only 10\% of the nominal service rate capacity in each container, which increases the use of the cloud (= higher cost) compared to bigger values of $\beta$.
As can be expected, all curves tend asymptotically to a minimum cost, which is that incurred when serving all the $\lambda$-apps on far-edge nodes.

\myfigeps{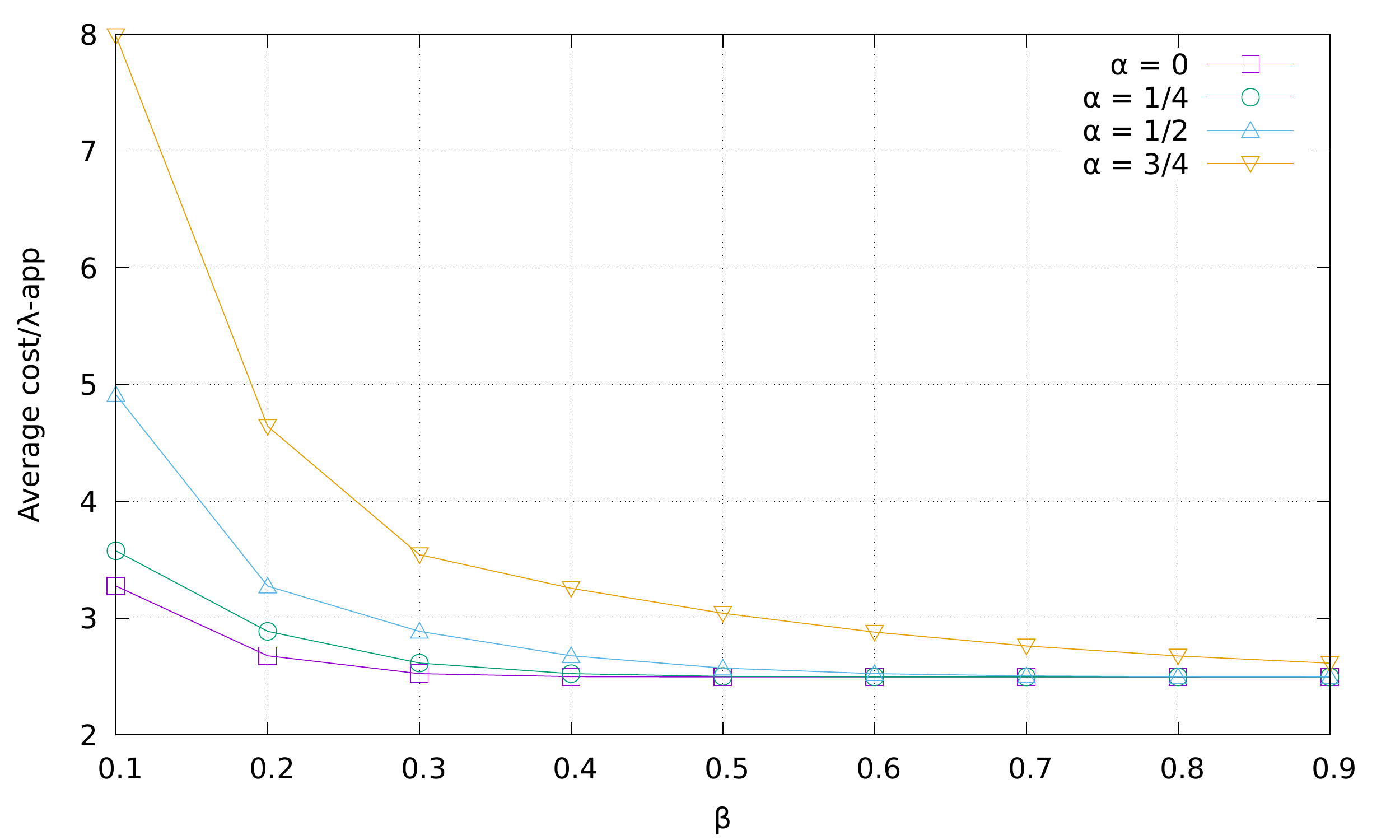}{%
Average unitary cost of $\lambda$-apps vs.\ $\beta$ for various values of $\alpha$, with $E[|A^\lambda|] = E[|A^\mu|]=50$.}

\myfigeps{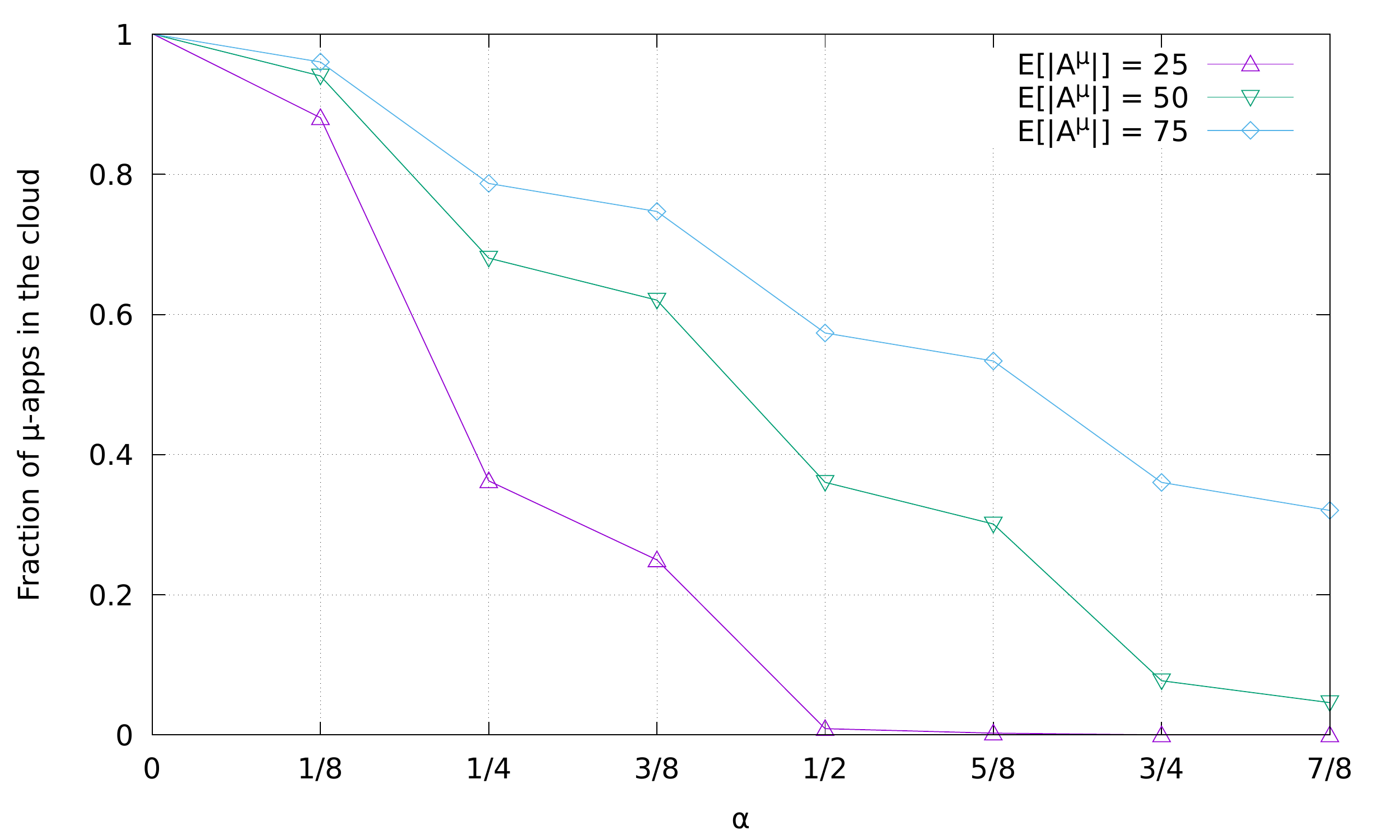}{%
Number of $\mu$-apps assigned to the cloud vs.\ $\alpha$.}

On the other hand, the impact of $\alpha$ on $\mu$-apps is shown in \cref{fig:010-mu-cloud}, in terms of the fraction of $\mu$-apps that are assigned a dedicated container in the cloud, hence at a higher cost in our model:
the curves with all values of $E[|A^\mu|]$ decrease almost linearly. 
The results confirm in a quantitative manner the intuition that the choice of $\alpha$ creates a trade-off between the performance of $\mu$-apps (better with bigger $\alpha$) and $\lambda$-apps (better with smaller $\alpha$), which leads the way to an auto-tuning of this parameter based on long-term optimization objectives, which is left for future work.


\subsection{Dynamic simulations}\label{sec:eval:dynamic}

In the dynamic simulations we use the Microsoft Azure traces described in \cref{sec:motivation} to drive an event-driven simulation of i) the clients, whose applications alternate over time between the $\mu$ and $\lambda$ operation modes according to the pattern that minimizes the respective user cost, and ii) the orchestrator, which performs both periodic optimization in \cref{sec:model:solution} and the best-effort measures reported in \cref{sec:model:operation}.
For simplicity of analysis, we set $\alpha = \beta = 0.5$.
Each simulation replication lasts 24~hours of simulated time and we run 6400 replications for each epoch duration, which increases from 1~minute to 30~minutes (we discard the samples in the first epoch as warm-up period to reduce initial bias effects).
The workload is composed of a number of applications drawn from a Poisson distribution with average $E[|A|] \in \{ 50, \ldots, 250 \}$, each assigned to a random broker in the network and exhibiting a random operation mode pattern from the traces, with randomized initial offset and wrap-around at the trace's end.

\myfigeps{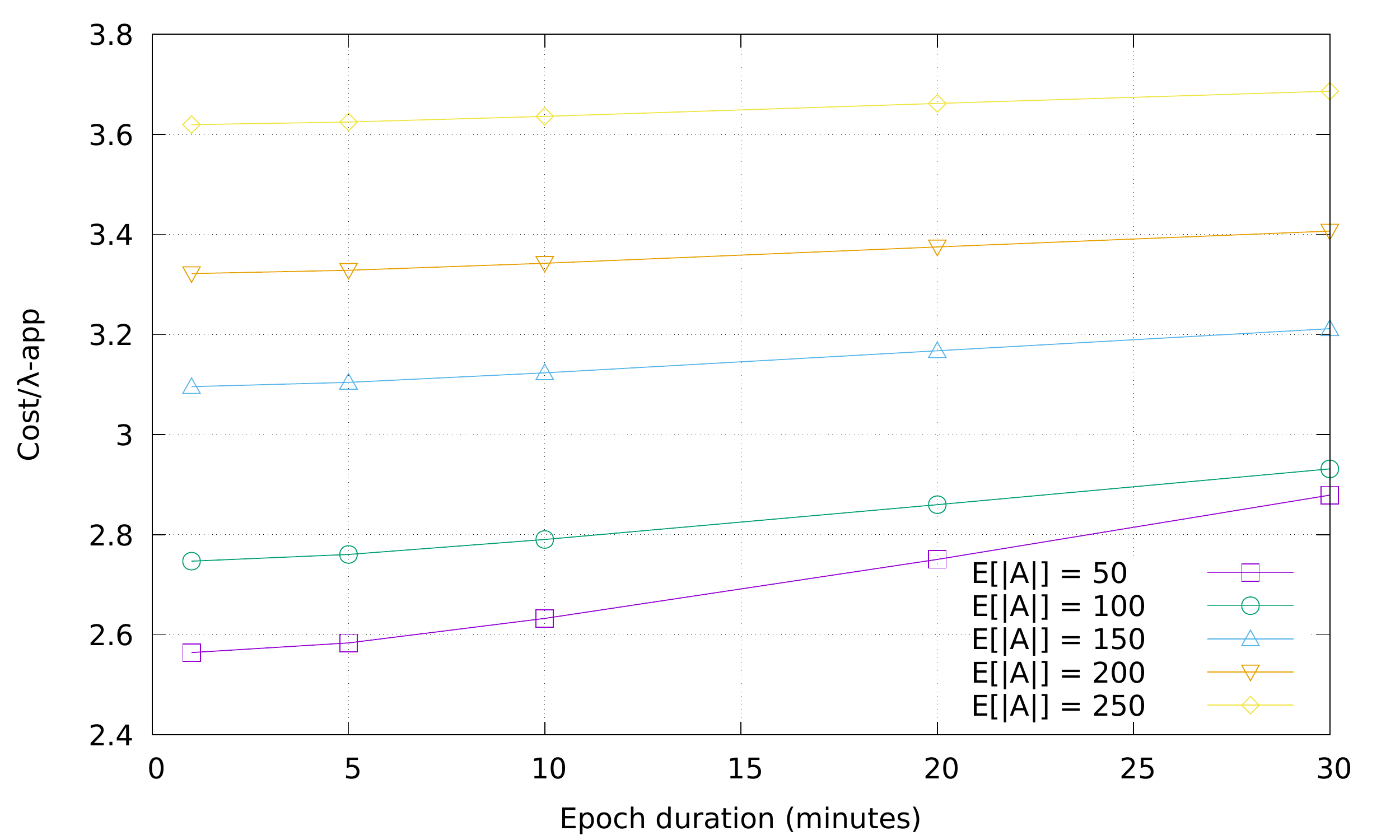}{%
Average cost of $\lambda$-apps vs.\ epoch duration for different workloads.}

As can be seen in \cref{fig:011-lambda-cost}, for all workloads, the unitary cost of $\lambda$-apps increases with the epoch duration.
This is because with larger optimization periods, there are (on average) more asynchronous transitions to $\lambda$ operation mode, which triggers the best-effort procedure in \cref{sec:model:operation} that merely directs them to the cloud.
The increase is more prominent with lighter workloads (i.e., $E[|A|]=50,100$) where the cloud is used more sparingly when solving the $\lambda$-apps allocation.

\myfigeps{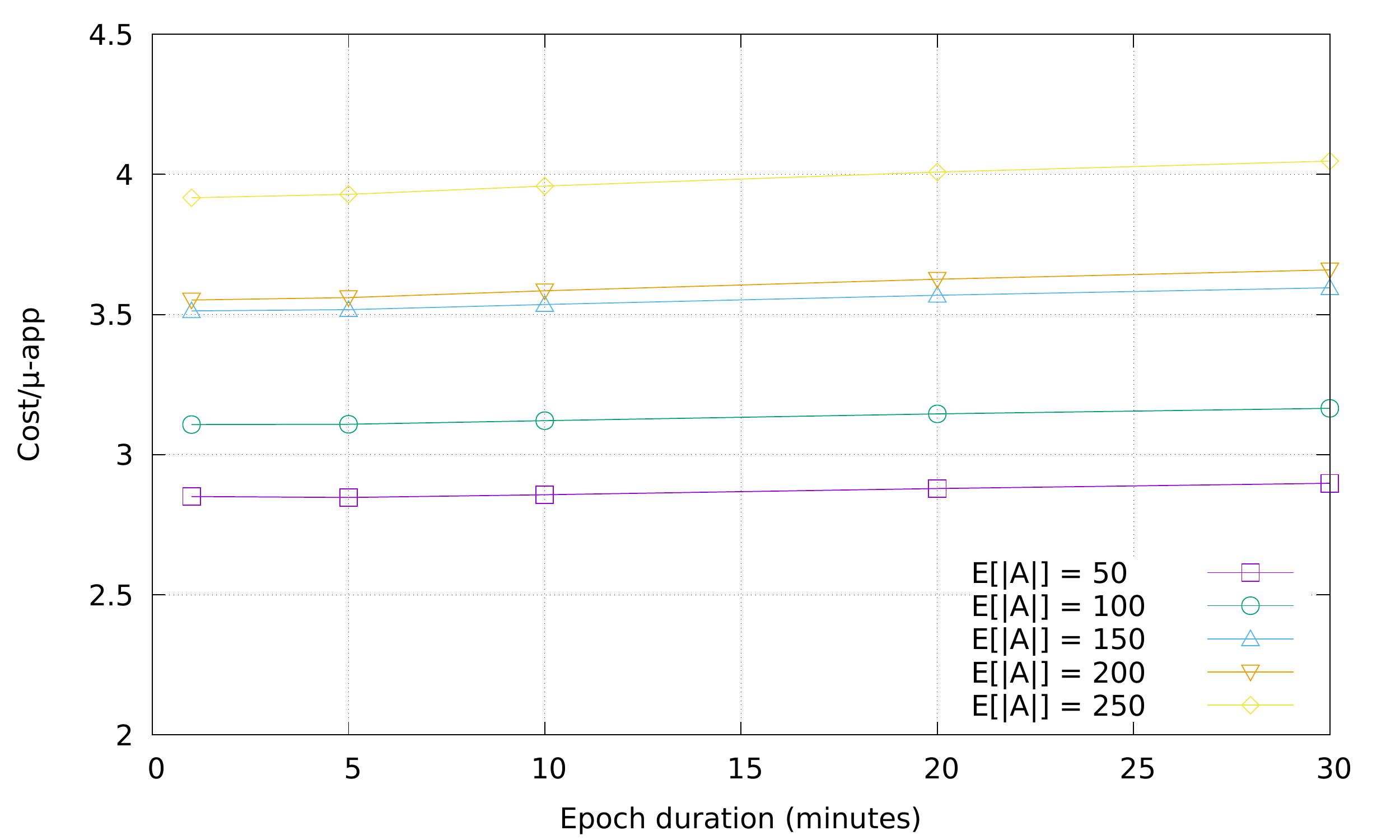}{%
Average cost of $\mu$-apps vs.\ epoch duration for different workloads.}

The cost of $\mu$-apps also increases with the epoch duration (see \cref{fig:011-mu-cost}), but only slightly because the best-effort procedure in \cref{sec:model:operation} recycles containers that are currently available for $\mu$-apps.
In \cref{fig:011-mu-cost-dist} we report the cumulative distribution over all the replications of the unitary cost of $\mu$-apps, for the representative case of 1 minute epoch duration.
As expected, the cost increases with the workload, but it is interesting to note that from $E[|A|] = 100$ to $E[|A|] = 150$ there is a wider gap, which happens because that is precisely when the orchestrator begins to use the cloud due to a shortage of edge resources for $\mu$-apps.
For a similar reason the curves $E[|A|] = 150$ and $E[|A|] = 200$ are almost overlapping: in that area, the marginal cost of adding more $\mu$-apps is small because most of them are still served by edge nodes.

\myfigeps{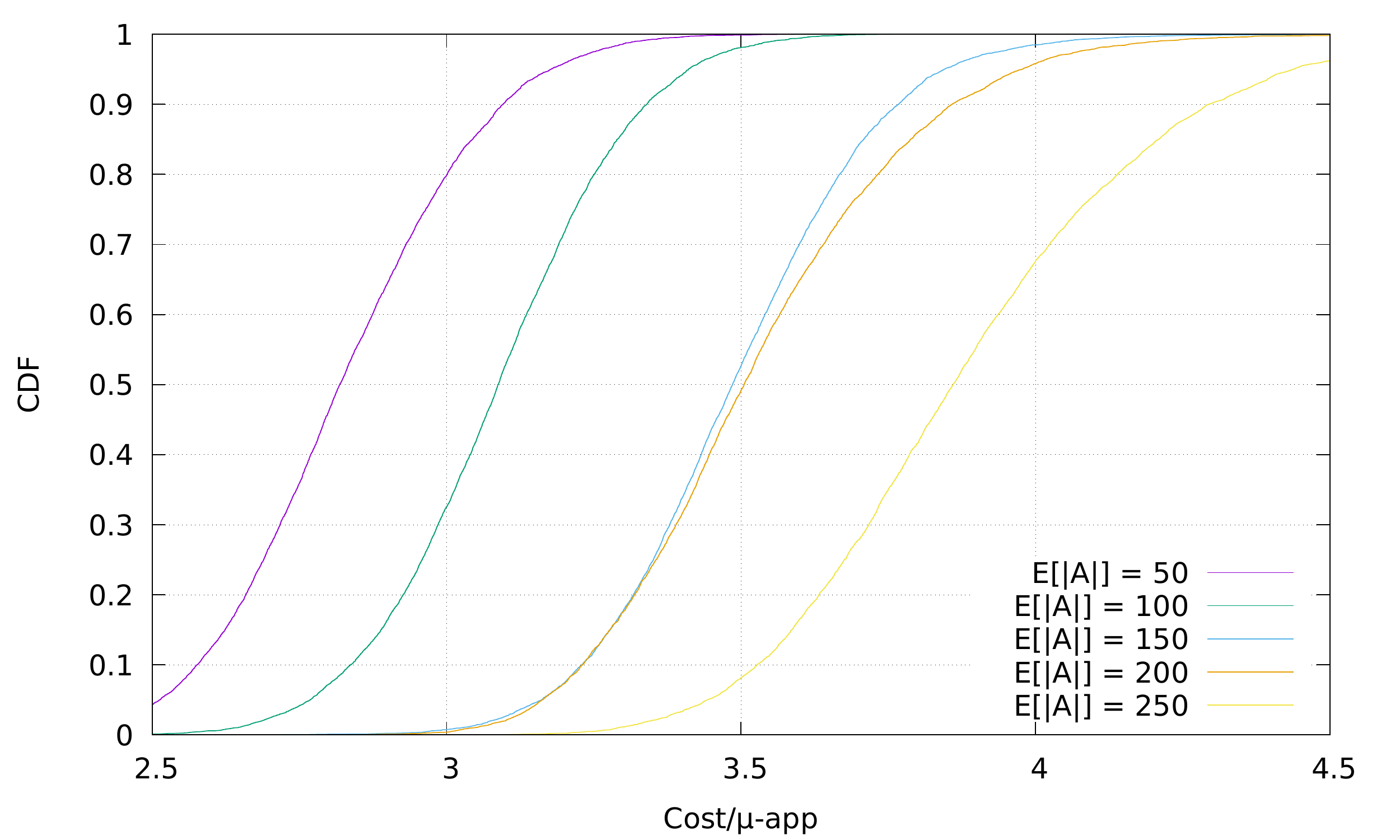}{%
Distribution of the cost of $\mu$-apps with 1 minute epoch duration.}

\myfigeps{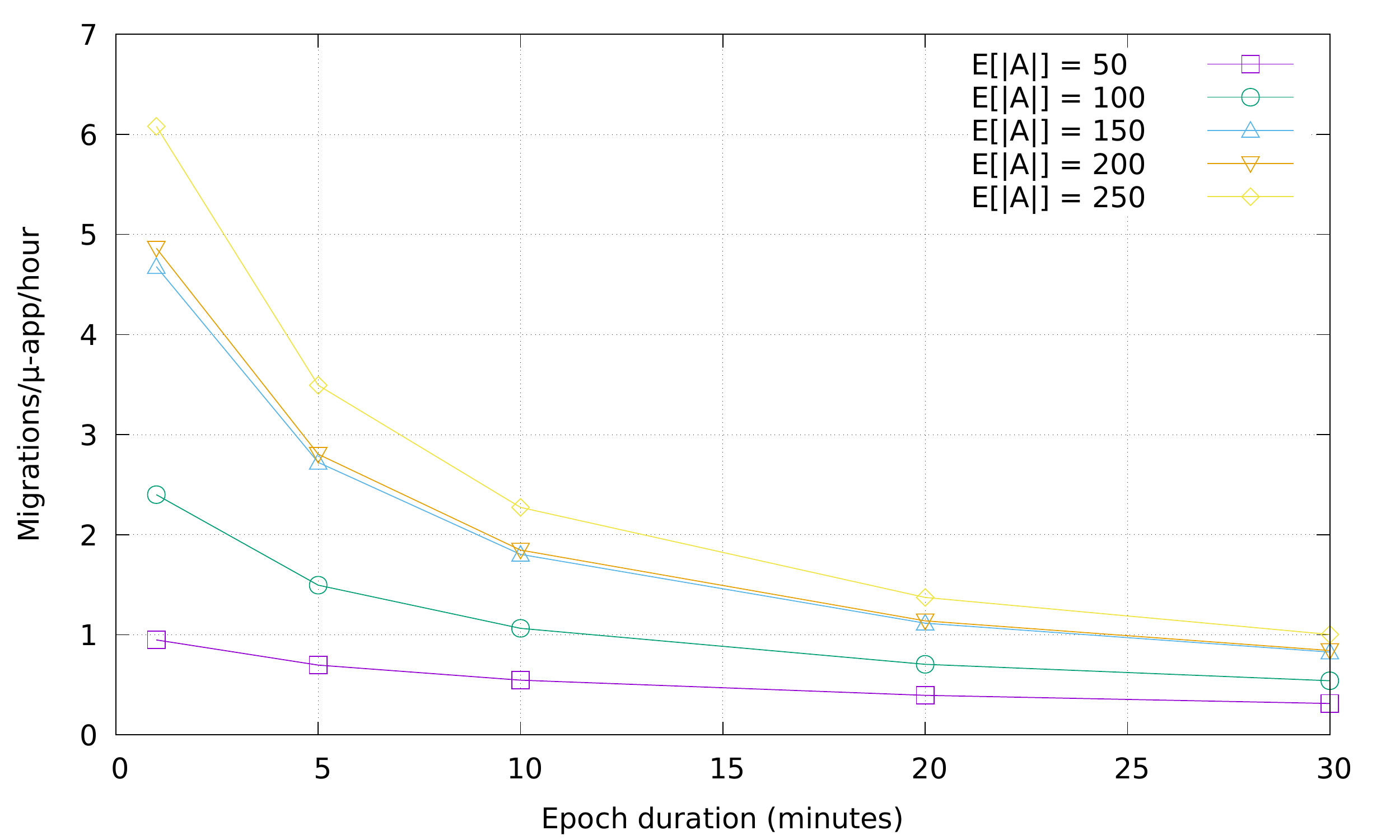}{%
Hourly rate of migrations vs.\ epoch duration for different workloads.}

We conclude with \cref{fig:011-mu-migrations}, which shows the migration rate of $\mu$-apps caused by executing the periodic optimization.
As can be seen, all the curves decrease significantly with the epoch duration.
This suggests that this system parameter should be set to a large value, which also reduces the rate of execution of the optimizations, hence the computational burden on the orchestrator.
However, this indication is in contrast with the cost analysis of both $\lambda$- and $\mu$-apps, thus a fundamental trade-off exists, which we plan to investigate more deeply in our future work, under realistic migration overhead costs.



\section{Conclusions}\label{sec:conclusions}

In this paper we have taken a novel perspective on the schism between the stateless ($\lambda$) and stateful ($\mu$) operation modes for edge-cloud applications.
In particular, based on the analysis of publicly available traces collected in a Microsoft Azure production environment, we have found that applications can benefit, in terms of operation costs, from alternating between the two operation modes over time.
This observation has led us to the definition of a mixed integer linear problem that jointly optimizes the resource allocation, in terms of containers assigned to $\mu$-apps and load distribution for $\lambda$-apps, depending on the instantaneous mode preferred by each application.
We have formulated the problem so that it can be solved efficiently in two sequential steps, meant to be executed periodically, and we have proposed best-effort measures to handle asynchronous changes in between consecutive optimization runs.
We have evaluated the performance of the proposed solution through comprehensive simulation experiments on a synthetic, but realistic, edge network topology and using a trace-driven workload composition.
The results have shown that our framework is flexible enough to adapt to a wide set of scenarios through the configuration of system parameters, including: the fraction of containers reserved for $\lambda$-apps ($1-\alpha$), the over-provisioning factor to absorb the peaks of $\lambda$-apps ($\beta$), and the epoch duration.
In our future work we will investigate how to set dynamically these parameters to achieve long-term optimisation objectives, we will study the impact of realistic migration overheads, and we will analyze the management plane protocols and programming interfaces for a practical implementation.

\section*{Acknowledgment}

This work was partially supported by the European Union's Horizon 2020 research and innovation programme under grant agreement No 957337, project \href{https://www.marvel-project.eu/}{MARVEL}, and by the Italian Ministry of Education
and Research (MIUR) in the framework of the CrossLab project (Departments of Excellence).

\balance



\end{document}